\documentclass[prl,twocolumn,amsmath,amssymb,showpacs]{revtex4}
\usepackage{graphicx}
\usepackage{dcolumn}
\usepackage{bm}
\begin{document}
\newcommand{\widthfigA}{0.35\textwidth} 
%
%
%
\title{From Lyapunov modes to the exponents for hard disk systems} 
\author{Tony Chung, Daniel Truant and Gary P. Morriss}
\affiliation{School of Physics, University of New South Wales, Sydney, 
New South Wales 2052, Australia}
\date{\today}
\begin{abstract}
   We demonstrate the preservation of the Lyapunov modes 
by the underlying tangent space dynamics of hard disks.
   This result is exact for the zero modes and correct to
order $\epsilon$ for the transverse and LP modes where
$\epsilon$ is linear in the mode number.
   For sufficiently large mode numbers the dynamics
no longer preserves the mode structure.
   We propose a Gram-Schmidt procedure based on orthogonality  
with respect to the centre space that determines the values of the 
Lyapunov exponents for the modes. 
   This assumes a detailed knowledge of the modes, but from that 
predicts the values of the exponents from the modes.
   Thus the modes and the exponents contain the same information.
\end{abstract}
\pacs{
05.45.Jn, 
05.45.Pq, 
02.70.Ns, 
05.20.Jj 
}
\maketitle  
%
%
%
   In a chaotic system, the difference between two nearby 
phase space trajectories, the so called Lyapunov vector, 
diverges exponentially in time. 
   If one or more of the rates of divergence are positive then 
the dynamics of a single initial condition is unpredictable 
and global behavior becomes important. 
   The statistical mechanics of chaotic many particle systems
is an example of a probabilistic treatment of the global
behavior of deterministic microscopic dynamics. 
   It is believed that the probabilistic axioms of statistical 
mechanics can be justified by the chaotic nature of the underlying
dynamics, so much effort has been devoted 
to finding links between macroscopic quantities, 
such as transport coefficients, 
and chaotic properties such as the Lyapunov exponents 
\cite{Gas98,Dor99,EM08}. 
   There have been some successes such as the conjugate 
pairing rule for the Lyapunov spectrum in some thermostated
systems \cite{ECM90,DM96,TM02b}
and the fluctuation theorem \cite{ECM93,GC95}.
   The stepwise structure of the Lyapunov spectrum 
(the full set of Lyapunov exponents for the system) is 
another chaotic property of many-particle systems 
which has been studied extensively 
\cite{Pos00,Mcn01b,Mil02,TM03a, TM03b, For04,Yan05,EF05,TM05a,TM05b,TM06b,TM07a}. 
   Each step in the exponent spectrum is associated with delocalised wavelike 
structures in the corresponding Lyapunov vector, 
now referred to as a Lyapunov mode.  

   The significance of the step structure is that it 
appears in the Lyapunov exponents which are closest to zero, 
connecting them with the slowest macroscopic modes of the system. 
   Some analytical approaches, such as random matrix 
theory \cite{Eck00,TM02a}, kinetic theory 
\cite{Mcn01b,Wij04} and periodic orbit theory \cite{TDM02} 
have been used in an attempt to understand this phenomenon. 
   A clue to understanding the stepwise structure of the Lyapunov 
spectrum and the Lyapunov modes is in the behavior of the Lyapunov 
vectors associated with the zero-Lyapunov exponents \cite{TM05b}. 
  In each case the Lyapunov modes associated with a zero exponent
are Noether transformations \cite{RM08} generated by the conserved quantities or 
time translation along the phase space trajectory.
  The Lyapunov modes associated with the steps
in the spectrum are $k$-vector analogues of the zero modes
and have the same basis in the fundamental symmetries of the
system \cite{TM05a,EF05,TM05b}.

   We consider a quasi-one-dimensional system (QOD) \cite{TM03a}
where the two-dimensional rectangular system is so narrow that the 
particles remain ordered in the $x$-direction. 
   We use Hard-wall boundary conditions 
in the $x$ direction and Periodic boundary conditions 
in the $y$ direction (the (H,P) boundary conditions).
   The step structure of the Lyapunov spectrum consists of one-point steps
and two-point steps  \cite{TM03a}. 
   The one-point steps correspond to transverse modes while
the two-points steps correspond to longitudinal and momentum
proportional modes. 
   The significant advantage of the QOD system is that both 
the exponents and the modes can be obtained to high accuracy
by standard numerical schemes \cite{Ben76,Ben80a,Shi79,
GPTCLP07,Sz07,Pazo08,YR08}.
   For systems with smooth interaction potentials it has proved
much more difficult to get clear numerical evidence for the steps
in the Lyapunov spectrum and to find modes \cite{For05,Yan05}. 
   However, the same structure must exist as the dynamics
is subject to the same invariances and conservation properties. 

   The QOD system with (H,P) boundary conditions we studied 
contained $N$ hard disk particles. 
   The complete description of the system at any time is contained
in the $4N$-dimensional phase vector $(q,p)$ where $q \equiv {\bf q}_{1},...{\bf q}_{N}$
and $p \equiv {\bf p}_{1},...{\bf p}_{N}$ where ${\bf q}_{i}$ and $ {\bf p}_{i}$ contain
the $x$ and $y$ coordinates and momentum of particle $i$.
   The time evolution of the phase vector through a free flight, and 
then collision, is given by the matrix equation
\begin{equation}\label{G_space}
\left(
\begin{array}{c}
q^{'}  \\
p^{'} 
\end{array}
\right)
=
\left(
\begin{array}{cc}
 \mathcal{I}  &  \tau \mathcal{I}   \\
 \mathcal{O}  &  \mathcal{N}
\end{array}
\right)
\left(
\begin{array}{c}
 q \\
 p
\end{array}
\right)
\end{equation}
where each scripted matrix is an $N \times N$ matrix containing $2 \times 2$ sub matrices,
$\mathcal{I}$ is the identity and $\mathcal{O}$ is the zero matrix.
  The matrix $\mathcal{N}$ that changes the momenta at collisions is given by
\begin{equation}\label{N_matrix}
\mathcal{N} = 
\left(
\begin{array}{cccccc}
I  & .  & 0 & 0  & . &  0   \\
.  & .  & .  & .  & .  &  .  \\
0  &  . &I-{\bf n}_{ij} {\bf n}_{ij}^{T}  & {\bf n}_{ij} {\bf n}_{ij}^{T}  &  .  & 0  \\
0  &  . &  {\bf n}_{ij} {\bf n}_{ij}^{T}  & I-{\bf n}_{ij} {\bf n}_{ij}^{T} & .  & 0  \\
.  & .  & .  & .  & .  &  . \\ 
0 & .  & 0  & 0  &  .  &  I  
\end{array}
\right).
\end{equation}
   Again, each element is itself a $2 \times 2$ sub-matrix.
   The term ${\bf n}_{ij}^{T} = \left(x_{ij}, y_{ij} \right)$ is a row vector containing the $x$ and $y$ components
of the separation between particles $i$ and $j$ at collision, so that the dyadic product 
${\bf n}_{ij}  {\bf n}_{ij}^{T} $ is a $2 \times 2$ matrix.


   We write the Lyapunov vectors as 
$(\delta q, \delta p)^{T}$ where $\delta q$ and  $\delta p$ 
are $N$-dimensional vectors containing the $2$-dimensional entries
for each particle position separation $\delta {\bf q}_{j}$ or momentum 
separation $\delta {\bf p}_{j}$. 
   For this QOD system there are four zero-Lyapunov exponents
and hence four associated Zero (Z) Lyapunov modes.
   The numerically observed Z modes can be written as linear 
combinations of elements of the basis set \cite{TM05b} 
$\delta \Gamma_{y} = (\delta q, 0)$, $\delta \Gamma_{py} = (0, \delta p)$, 
$\delta \Gamma_{t} =( p,  0)/||p||$ and  $\delta \Gamma_{e} =(0,  p)/||p|| $
where the $j^{th}$ element of $\delta q$ and $\delta p$ 
in $\delta \Gamma_{y}$ and $\delta \Gamma_{py}$ is given by
$\delta {\bf q}_{j} \sim \delta {\bf p}_{j} \sim \frac {1} {\sqrt{N}}  \left( \begin{array}{c} 0\\
1 \end{array} \right)$, 
while $p$ is an $N$-dimensional vector whose $j^{th}$ component is 
${\bf p}_{j}$ (the momentum of particle $j$)   
and $||p||$ is the magnitude of the total momentum. 


   The time evolution equations for tangent space dynamics consist of many repeats 
of a free-flight followed by a collision and the Gram-Schmidt procedure. 
   The first two of these steps, the application of a free-flight matrix then a collision  
matrix, evolve the tangent vector in time from $0$ to $\tau$ and can be written as
\begin{equation}\label{dG_space}
\left(
\begin{array}{c}
\delta q (\tau) \\
\delta p (\tau)
\end{array}
\right)
=
\left(
\begin{array}{cc}
 \mathcal{N}  &  \mathcal{O}   \\
 \mathcal{Q}  &  \mathcal{N}
\end{array}
\right)
\left(
\begin{array}{cc}
 \mathcal{I} &  \tau \mathcal{I}   \\
 0  &  \mathcal{I}
\end{array}
\right)
\left(
\begin{array}{c}
\delta q \\
\delta p
\end{array}
\right)
\end{equation}
where 
\begin{equation}\label{Q_matrix}
\mathcal{Q} = 
\left(
\begin{array}{cccccc}
0  & .  & 0 & 0  & . &  0   \\
.  & .  & .  & .  & .  &  .  \\
0  &  . &-Q_{ij}  & Q_{ij}  &  .  & 0  \\
0  &  . &  Q_{ij}  & -Q_{ij}  & .  & 0  \\
.  & .  & .  & .  & .  &  . \\
0 & .  & 0  & 0  &  .  &  0  
\end{array}
\right).
\end{equation}
   
   Each component of the matrix $\mathcal{Q}$  is a $2\times 2$ sub-matrix where 
the only non-trivial components are those associated with the two particles that collide, $i$
and $j$ through $Q_{ij}$.
   Then $Q_{ij}$ given is by 
\begin{equation} \label{Qij}  
   Q_{ij}=({\bf n}_{ij} \cdot {\bf p}_{ij}) \left[I+\frac {{\bf n}_{ij} {\bf p}^{T}_{ij}} {{\bf n}_{ij} \cdot {\bf p}_{ij}} \right] \cdot  \left[ I- \frac {{\bf p}_{ij} {\bf n}^{T}_{ij}} {{\bf n}_{ij} \cdot {\bf p}_{ij}} \right].
\end{equation}
where ${\bf p}_{ij}^{T} = \left(p_{xij}, p_{yij} \right)$ is a row vector containing the $x$ and $y$ 
components of the relative momenta at collision (${\bf p}_{ij} = {\bf p}_{j} - {\bf p}_{i}$).
   The principle property that we will exploit now is that $Q_{ij} \cdot {\bf p}_{ij} = 0$.

  To understand the tangent vector dynamics we consider the action of the matrix 
$\mathcal{N}$ on either $\delta q$ or $\delta p$ which gives
\begin{equation}\label{N_action}
\mathcal{N} \delta q = \delta q + {\bf n}_{ij} {\bf n}_{ij}^{T} \cdot (\delta {\bf q}_{j} - \delta {\bf q}_{i}) 
X.
\end{equation}
where $X$ is the $N$-dimensional column vector with all elements equal to zero 
except for $X_{i}=1$ and $X_{j}=-1$.
   Note that for a QOD system $j=i+1$, but otherwise the result is general.
   
   Similarly, the action of the matrix $\mathcal{Q} $ qives
\begin{equation} \label{Q_action}
\mathcal{Q} \delta q = Q_{ij} \cdot (\delta {\bf q}_{j} - \delta {\bf q}_{i})
X
\end{equation}


   Any Lyapunov mode for which $\delta {\bf q}_{i} - \delta {\bf q}_{j} = 0$ or 
$\delta {\bf p}_{i} - \delta {\bf p}_{j} = 0$ exactly - such as the zero
mode $\delta \Gamma_{y}$ - is preserved by the dynamics,
while the conjugate zero mode grows linearly with time, so for example  
$\delta \Gamma_{py} (\tau) =
 \delta \Gamma_{py} + \tau \delta \Gamma_{y}$ \cite{RM08}.

   
   The application of the dynamics on the momentum dependent $Z$ modes
gives  for $\delta \Gamma_{t}$ that $\mathcal{N} p=p^{'}$
and $\mathcal{Q} p=0$ so that $( p, 0)/||p|| \rightarrow ( p^{'},  0)/||p^{'}||$ which is 
$\delta \Gamma_{t}$ at the new time.
   Therefore the functional form of $\delta \Gamma_{t}$ is preserved exactly.
   The conjugate mode is linear in time $\delta \Gamma_{e} (\tau) =
 \delta \Gamma_{e} + \tau \delta \Gamma_{t}$.   

   The numerically observed Lyapunov modes are of four types, 
$Z$ modes, transverse ($T$) modes, longitudinal ($L$) modes
and momentum proportional ($P$) modes.
  The $L$ and $P$ are usually observed together as a combined $LP$ mode. 
  The $T$ modes are given by
\begin{equation}\label{TL_modes}
\delta T^{n} =
\left( \begin{array}{c}
\delta q \\
\delta p
\end{array} \right)
=
\left( \begin{array}{c}
\gamma_{n} \delta q_{T}\\
\gamma^{'}_{n}  \delta p_{T} 
\end{array} \right)
\end{equation}
where $\gamma$ and $\gamma^{'}$ are constants \cite{MT09}.
   The components of both $\delta q_{T}$ and $\delta p_{T}$
are of the form $\delta {\bf q}_{j} = \left( \begin{array}{c} 0\\
c_{nj} \end{array} \right)$ where $c_{nj} = \cos k_{n} x_{j}$.


   In a transverse mode the term $(\delta {\bf q}_{j}- \delta {\bf q}_{i})$
in equations (\ref{N_action}) and (\ref{Q_action}), 
becomes $(\delta y_{j}-\delta y_{i})=- \epsilon \gamma_{n} \sin(k_{n} x_{i})$ 
where $\epsilon = k_{n} x_{ij} = n \pi x_{ij}/L$.
   Here $x_{ij}$ is the $x$-component of the distance between particles
$i$ and $j$ at collision and $L$ is the length of the system in the $x$-direction.
   The small parameter $\epsilon$ is linear in the mode number $n$. 
      Thus the time evolution of a $T$ modes is given by
\begin{equation}\label{mode_evol}
\left( \begin{array}{c}
\delta q(\tau) \\
\delta p(\tau)
\end{array} \right)
=
\left( \begin{array}{c}
\delta q + \tau \delta p\\
\delta p
\end{array} \right)
+ O (\epsilon).
\end{equation}
and henceforth we neglect the $\epsilon$ dependent terms. 
   Clearly for sufficiently large $\epsilon$ the dynamics without
the order  $\epsilon$ term will become incorrect.
   We assume that below a threshold value of $\epsilon$ the
dynamics in equation (\ref{mode_evol}) is correct.


   Further, for the $LP$ mode, the $P$ component is linear in the momentum, 
so that for example 
$\delta {\bf q}_{i} = \beta_{n} {\bf p}_{i} \cos k_{n} x_{i}$ then 
$\delta {\bf q}_{i} - \delta {\bf q}_{j} =\beta_{n}({\bf p}_{ij} \cos k_{n} x_{i} - 
\epsilon {\bf p}_{j} \sin k_{n} x_{i})$.
   As $Q_{ij} \cdot {\bf p}_{ij} = 0$ the first term is zero and the result is order $\epsilon$.
   The $LP$ modes are of the form
\begin{equation}\label{LP_modes}
\delta LP^{n}_{1} =
\left( \begin{array}{c}
\delta q \\
\delta p
\end{array} \right)
= s_{nt}
\left( \begin{array}{c}
\beta_{n} p c \\
\beta^{'}_{n}  p c
\end{array} \right)
+ c_{nt}  \left( \begin{array}{c}
\alpha_{n} \delta q_{L} \\
\alpha^{'}_{n}  \delta p_{L}
\end{array} \right)
\end{equation}
where $\beta_{n}$, $\beta^{'}_{n}$, $\alpha_{n}$ and $\alpha^{'}_{n}$ are constants \cite{MT09}.  
   The $N$-dimensional vector $p c$ has components ${\bf p}_{j} c_{nj}$,
$s_{nt} = \sin \omega_{n} t$ 
and $c_{nt} = \cos \omega_{n} t$ (with frequency $\omega_{n}$). 
   In the case of (H,P) boundary conditions the $LP$ exponents are doubly degenerate so
there is a second $LP$ mode $\delta LP^{n}_{2}$ 
with $s_{nt}$ and $c_{nt}$ interchanged which is orthogonal in time to $\delta LP^{n}_{1}$.
   This result means that the $LP$ modes also have a time evolution governed
by equation (\ref{mode_evol}) with a different order $\epsilon$ term and again 
we assume that this dynamics is correct below some threshold $\epsilon$.


   The numerically observed $T$ modes are invariants of the dynamics 
and the pair of $LP$ modes define a two-dimensional sub-space.
   The action of the dynamics and Gram-Schmidt procedure therefore 
simply results in the $T$ mode remaining 
orthogonal to the centre space defined by the zero modes.
   This suggests that we can define an {\it inside-out} Gram-Schmidt procedure
to obtain the same effect.
   To do this we make two assumptions:
   1) that the functional forms for the $T$, $L$ and $P$ modes are known;
   2) that the numerical values for the coefficients are known.


   Given these assumptions we can calculate the values of the Lyapunov exponents
in the step region. 
   If we consider a $T$ or $L$ mode then under
a free flight and collision using equation (\ref{mode_evol}) the mode changes, 
but then the Gram-Schmidt procedure returns it to its initial direction with
some scaling factor $\zeta$, as   
\begin{equation}\label{t_evol_gs} 
\left(
\begin{array}{c}
\delta q  \\
\delta p 
\end{array}
\right)
\stackrel  {\tau,coll} {\rightarrow} %
\left(
\begin{array}{c}
\delta q + \tau \delta p   \\
 \delta p
\end{array}
\right)
\stackrel {GS} {\rightarrow} 
\zeta
\left(
\begin{array}{c}
\delta q \\
\delta p
\end{array}
\right)
\end{equation}
   The first right arrow is the action of the free flight and collision while the
second right arrow is the result of the Gram-Schmidt procedure. 
   Here we use a simplified {\it inside-out} Gram-Schmidt procedure in which
we assume that the mode is already orthogonal to the centre space and 
then ensure orthogonality with respect to
the conjugate mode.


   Using the symplectic property of the system \cite{TM03b} the mode conjugate to $\delta T^{(n)}$  is
\begin{equation}\label{conjugate_mode}
\delta T^{(-n)} = 
\left(
\begin{array}{c}
 - \delta p   \\
 \delta q 
\end{array}
\right).
\end{equation}
   Applying the {\it inside-out} Gram-Schmidt procedure to equations (\ref{t_evol_gs}) 
gives
\begin{equation}\label{zeta}
\zeta 
\left(
\begin{array}{c}
\delta q \\
\delta p
\end{array}
\right)
= \left(
\begin{array}{c}
\delta q + \tau \delta p \\
\delta p
\end{array}
\right)
+ \tau (\delta p \cdot \delta p)
\left(
\begin{array}{c}
-\delta p \\
\delta q
\end{array}
\right)
\end{equation}
or two equations to solve for the scale factor $\zeta$.
   It is straigthforward to see that as the mode is normalised
$\delta q \cdot \delta q + \delta p \cdot \delta p = 1$  so both
components of equation (\ref{zeta}) give the same solution
\begin{equation}\label{soln_alpha}
\zeta = 1 + \tau \frac {(\delta p \cdot \delta p) (\delta q \cdot \delta q)} {\delta q \cdot \delta p}
\end{equation}
   The full time evolution is infinitely many repeats of this process: free-flight,
collision and Gram-Schmidt, so the Lyapunov exponent is given by 
\begin{equation}
\lambda = \lim_{m \rightarrow \infty} \frac {1} {T} \ln \prod_{i=1}^{m} \zeta_{i} 
\end{equation}


   For the mode $\delta T^{n} $, $\delta y_{j} = \gamma_{n} c_{nj}$
and $\delta p_{yj} = \gamma^{'}_{n} c_{nj}$ so assuming that
$\sum_{j}^{N} c_{nj}^{2} = N/2$, we have
\begin{equation}\label{expT}
\lambda_{n} = \frac {N} {2} \gamma^{'}_{n} \gamma_{n}.
\end{equation}
   Similarly, we can consider the evolution of the negative mode $\delta T^{-n}$ and by 
ensuring orthogonality with respect to its conjugate mode $\delta T^{n} $,
the result will be  $\lambda_{-n} = \frac {N} {2} \gamma^{'}_{-n} \gamma_{-n}
= -\lambda_{n}$.
 

   Next treat the longitudinal part of the $LP$ mode without its
explicit time dependence.
   Clearly this is just the same as the transverse mode and the result is
\begin{equation}\label{expL}
\lambda_{n} = \frac {N} {2} \alpha^{'}_{n} \alpha_{n}.
\end{equation}


   We treat the momentum dependent part of the $LP$ mode without
its explicit time dependence and ensure orthogonality with respect to
the conjugate $P$ component.
   A similar argument leads to the result
\begin{equation}\label{expP}
\lambda_{n} =N  \beta^{'}_{n} \beta_{n} T.
\end{equation}
where the temperature is given by $2NT = \sum_{j} {\bf p}_{j}^{2}$.   
  Again the negative exponent is simply $\lambda_{-n} = -\lambda_{n}$
for both $L$ and $P$ components of the mode.



\begin{table}[t]
\caption{
      A comparison of predicted (eqns \ref{expT}, \ref{expL}, \ref{expP}) and numerically observed
      Lyapunov exponents for a $200$ particle QOD system
      with (H,P) boundary conditions at density $\rho=0.8$ and temperature $T=1$. 
      We use units where the mass and disk radius $R$ are $1$, the total energy is $N$ 
    and the system size is $L_{y} = 1.15 R$ and $L_{x} = N/\rho L_{y}$ 
    where $\rho$ is the density. There are small differences between the $x$ and $y$
    projections of $\beta$ so we include both results.
   }
\vspace{0.2cm}
\begin{tabular}{ccc|cccc} \hline\hline
  $n$
    & $\frac {N} {2} \gamma \gamma^{'} $
    & $\lambda_{T}$ 
    & $ N \beta_{x} \beta^{'}_{x} T$
    & $ N \beta_{y} \beta^{'}_{y} T$
    & $\frac {N} {2} \alpha \alpha^{'} $
    & $\lambda_{LP}$ 
    \\ \hline
1 & 0.0388 & 0.0393 & 0.0599 & 0.0559 & 0.0502 & 0.0605  \\
2 & 0.0749 & 0.0784 & 0.1169 & 0.1044 & 0.0965& 0.1229 \\
3 & 0.1099 & 0.1177 & 0.1507 & 0.1348 & 0.1344 & 0.1848\\
4 & 0.1431 & 0.1571 & 0.1753 & 0.1472 & 0.1591 & 0.2484\\
5 & 0.1748 & 0.1961 & 0.1693 & 0.1391 & 0.1595 & 0.3140 \\
6 & 0.2007 & 0.2352 & 0.1855 & 0.1409 & 0.1823 & 0.3791 \\
\hline\hline
\end{tabular}
\label{compare_exp}
\end{table}
 

   In table \ref{compare_exp} we compare the predicted and numerical results
for the Lyapunov exponents of the first six modes of each type.
   The first $T$ mode and $P$ mode are quite accurate but the first $L$ mode is $20\%$
less than the numerical result.
   Generally the results become worse for higher order modes.

   There are a number of possible sources of error.
   The dynamics is limited by the size of the neglected term $\epsilon$ and will
be worse for larger $n$.
   Probably the most important limitation is the simplified {\it inside-out} Gram-Schmidt
procedure as it assumes that the functional forms for the modes are already 
orthogonal to the centre space of $Z$ modes.
   At any finite $N$ this is not correct and any more exact  {\it inside-out} Gram-Schmidt
procedure would need to work systematically ensuring orthogonality with all previous
modes.
   Thus, for example, to Gram-Schmidt the 3rd $T$ mode it should be explicitly made orthogonal to
the centre space, the 1st and 2nd $T$ modes and any $LP$ modes with lower value exponents.
    

   In conclusion, we have shown that the Lyapunov exponents for all types of modes 
can be calculated using an {\it inside-out} Gram-Schmidt procedure and a complete
knowledge of the functional form of the modes.
   The simplified Gram-Schmidt procedure is only accurate for the first T and P mode
components but the systematic approach suggested above may improve the accuracy
but at the cost of the simplicity of the result.
   Thus we see that the same information that is encoded in the modes is also
encoded in the values of the exponents.

   In all numerical calculations of Lyapunov modes there are a set of modes which are 
stable below some maximum mode number $n_{max}$.
   Here we require that $\epsilon$ in the dynamics of equation (\ref{mode_evol}) 
is less than some threshold $\epsilon_{max}$
which we can estimate from the numerics used to generate table \ref{compare_exp}.
   For a system of $N=200$ disks we find that $n_{max} \sim 24$, 
and for the QOD system $x_{ij}$ is positive and
bounded by $\sqrt {1 - L^{2}_{y} / 4 } < x_{ij} < 1$.
   Therefore $\epsilon_{max} \sim 0.35 x_{ij}$ which corresponds to about $8$
particles per half wavelength.
   If the initial Lyapunov vector is mode-like for a particular $n$ then the dynamics
will preserve its mode-like character for $n < n_{max}$ and it will be unstable for  $n > n_{max}$.
   The question of the stability or otherwise of a particular mode
is a different level of stability for this system.
%
%

%
\end{document}